\documentclass[%
 reprint,
 superscriptaddress,
 amsmath,amssymb,
 aps,
 prx,
 nofootinbib,
]{revtex4-2}

\usepackage{graphicx}    
\usepackage{booktabs}    
\usepackage{xcolor}      
\usepackage{microtype}   
\usepackage{tikz}        
\usetikzlibrary{arrows.meta, positioning, fit, calc}
\usepackage{multirow}    

\usepackage{hyperref}

\begin{document}

\preprint{APS/123-QED}

\title{Real-Time Plasma State Prediction via FPGA-Accelerated Quantized Recurrent Probabilistic Neural Networks}

\author{Daniel Gaytan-Villarreal}
\thanks{Corresponding author: jgaytanv@andrew.cmu.edu}
\affiliation{Department of Physics, Carnegie Mellon University, Pittsburgh, PA 15213, USA}

\author{Aiken Xie}
\affiliation{Department of Applied Physics and Applied Mathematics, Columbia University , New York, NY 10027, USA}

\author{Tu Pham}
\affiliation{Department of Electrical and Computer Engineering, Georgia Institute of Technology, Atlanta, GA 30332, USA}

\author{Rohit Sonker}
\affiliation{Machine Learning Department, Robotics Institute, Carnegie Mellon University, Pittsburgh, PA 15213, USA}


\author{Chiara Amendola}
\affiliation{Department of Physics, Carnegie Mellon University, Pittsburgh, PA 15213, USA}

\author{Matteo Cremonesi}
\affiliation{Department of Physics, Carnegie Mellon University, Pittsburgh, PA 15213, USA}

\author{Cong Hao}
\affiliation{Department of Electrical and Computer Engineering, Georgia Institute of Technology, Atlanta, GA 30332, USA}

\author{Jeff Schneider}
\affiliation{Machine Learning Department, Robotics Institute, Carnegie Mellon University, Pittsburgh, PA 15213, USA}

\date{\today}

\begin{abstract}
Real time plasma state estimation for control of Tokamak devices are challenging due to the stringent latency requirements of the plasma control system (PCS). We present an end-to-end workflow for deploying a recurrent probabilistic neural network
(RPNN) on FPGA hardware. We combine architecture size reduction with quantization-aware training via
QKeras. The model is then synthesized using hls4ml, targeting a
Xilinx Alveo U50 device. We report a design that fits comfortably within all four
resource budgets (DSP, LUT, FF, BRAM) at deterministic sub-10~$\mu$s single-timestep
latency, meeting the requirements for real-time inference inside a
model-predictive-control-style plasma control loop. 
\end{abstract}

\maketitle

\section{Introduction}

\subsection{Real-time State Estimation \\ for Tokamak Plasma Control}
\label{sec:motivation}

Tokamak plasmas exhibit highly complex, nonlinear dynamics. Modern tokamak plasma control systems (PCS) are increasingly moving beyond
independent single-quantity feedback loops towards integrated model-based control
strategies~\cite{humphreys2015novel}. Model-predictive-control
(MPC)-style approaches are particularly attractive for this task, but they require many lookaheads of the plasma state given actuator input within a single
control cycle, placing a hard latency budget on the plasma state evolution model.
Physics-based transport simulations are generally far too slow, which
has motivated growing interest in fast, data-driven surrogate dynamics models trained on large datasets of existing experimental discharges. However, such surrogates are only useful for real-time control, if they can be evaluated with low, deterministic latency. Such a requirement is difficult to satisfy with a model executed on a general-purpose processor, and instead points toward a dedicated hardware implementation. FPGA-hosted neural networks already serve such latency-critical roles across experimental physics, from real-time event selection at the Large Hadron Collider to qubit readout and control~\cite{govorkova2022autoencoders,deiana2022fastml,khoda2022ultralow,gaytan2026chargejump}.

A further challenge is that the implementation must capture the temporal structure of plasma evolution, which is inherently sequential: the state at a given time depends not only on the present measured state and actuator settings but on dynamics (e.g.,~current profile diffusion, rotation and density transport) that are not fully captured by an instantaneous snapshot of the plasma. Recurrent probabilistic neural networks (RPNNs) are a natural
architectural choice for this setting, as their hidden state can in principle carry forward exactly this kind of unobserved dynamical information across timesteps \cite{char2024fullshot}.

Deploying the RPNN inside a real-time control loop requires translating this trained model into deterministic, low-latency hardware. \textsc{hls4ml}~\cite{Duarte:2018ite} has become a standard workflow for translating trained neural networks into synthesizable high-level-synthesis (HLS) firmware for FPGA-based real-time trigger and control systems across physics experiments.

This paper details the workflow used to adapt the RPNN for FPGA deployment: describing the model architecture and training curriculum (Sec.~\ref{sec:algorithm}), reducing the architecture's size and quantizing it to low fixed-point precision (Sec.~\ref{sec:ablation-quant}), validating the accuracy of the resulting design (Sec.~\ref{sec:performance}), and characterizing the resulting hardware resource and latency trade-offs to obtain a design suitable for real-time deployment (Sec.~\ref{sec:firmware}).

\subsection{The Recurrent Probabilistic Neural Network Dynamics Model}

In this work, we consider a recurrent probabilistic neural network model~\cite{char2024fullshot}. The model predicts the next full plasma state from the current state, the current actuator settings, and the commanded changes in those settings over the following timestep. It combines a gated recurrent unit (GRU)~\cite{cho2014gru} with encoder and decoder multilayer perceptron (MLP) subnetworks. The plasma state is a 27-dimensional vector comprising scalar quantities such as normalized beta, line-averaged density, and internal inductance as well as principal-component coefficients of several one-dimensional plasma profiles, such as plasma density, temperature, etc.

The RPNN outputs both a mean next-state prediction and a per-channel predictive log-variance, giving an estimate of aleatoric uncertainty at every timestep. An ensemble of independently trained RPNN members, obtained via bootstrap resampling of the training discharges and independent weight initialization, is further used to estimate epistemic uncertainty, so that the combined ensemble output supports uncertainty-aware downstream control decisions in addition to a point estimate of the plasma trajectory. Dynamics models of this kind have recently been deployed for downstream control of tokamaks, e.g.,~offline reinforcement learning for rotation-profile control~\cite{sonker2026offline,char2023offline}.

\section{Algorithm Design and Training}
\label{sec:algorithm}

\subsection{Model Architecture}
\label{sec:architecture}

This work adopts the RPNN architecture and training curriculum of
Ref.~\cite{char2024fullshot}, with one modification: the sequence-normalization layer following the encoder, originally a LayerNorm, is replaced with BatchNorm (Batch Normalization) for compatibility with the QKeras/hls4ml quantization and synthesis pipeline. We validate later in this subsection that this substitution carries no accuracy cost on our dataset. This modified architecture, using BatchNorm throughout, is the baseline used for every result in this paper.

Table~\ref{tab:state-actuator} catalogs the state and actuator variables making up the RPNN's input and output. The state is a 27-dimensional vector combining 7 scalar plasma parameters with 20 Principal Component Analysis (PCA) coefficients spanning 6 one-dimensional profile quantities; the actuator vector is a 13-dimensional set of external control inputs. The top PCA components are selected such that $99\%$ of the total variance is explained. The model's full input at each timestep concatenates the current state, the current actuator settings, and the commanded delta in those actuator settings over the following timestep ($27+13+13=53$ dimensions), and predicts the next state (27 dimensions).

\begin{table*}[t]
    \caption{List of all state and actuator variables. The total dimension (40) counts
        each unique variable once; the model's actual per-timestep input is
        53-dimensional, since the 13 actuator channels are each supplied twice --- once
        as their current absolute level and once as the commanded delta (change) into
        the following timestep. See
        Fig.~\ref{fig:architecture-ablated}.}
    \label{tab:state-actuator}
    \begin{ruledtabular}
    \begin{tabular}{llp{9cm}r}
    \textrm{Group} & \textrm{Representation Type} & \textrm{Signal} & \textrm{Dimension} \\
    \colrule
    \multirow{2}{*}{States}
            & Scalar  & $\beta_N$, line-averaged density, internal inductance ($l_i$), $q_0$, $q_{95}$, $v_{\rm loop}$, WMHD & 7 \\
            & Profile & electron temperature, ion temperature, density, rotation (4 PCA components each); pressure, Safety factor $q$ (2 PCA components each) & 20 \\
    \colrule
    \multirow{5}{*}{Actuators}
            & Beam         & power injected, torque injected & 2 \\
            & Shape        & $a_{\rm minor}$, triangularity (top/bottom), $\kappa$, $R$ and $Z$ coordinates of magnetic axis & 6 \\
            & Gas          & gas A puffing & 1 \\
            & Other        & plasma current target, toroidal field (magnitude + sign) & 3 \\
            & Aux.\ heating & ECH power & 1 \\
    \colrule
    \multicolumn{3}{r}{Total Dimension:} & 40 \\
    \end{tabular}
    \end{ruledtabular}
\end{table*}

This state/actuator representation is populated from real DIII-D tokamak discharges produced with the data pipeline of
Ref.~\cite{char2024fullshot}. We use $19{,}648$ shots in total, each resampled to a uniform $20\,$ms cadence. This dataset retains each shot's ramp-up phase and flattop, so the prediction task spans both the strongly-driven transient and the comparatively stable flattop.
Following Ref.~\cite{char2024fullshot}'s split methodology, shots are sorted
chronologically and partitioned $90/5/5$: the $17{,}676$ oldest shots ($55{,}566$ training sequences of up to $280$ timesteps, counted from each shot's start) form the training set, the following $982$ shots ($2{,}407$ sequences) the validation set, and the $990$ most recent shots ($3{,}000$ sequences) the held-out test set. The three shot-number ranges are strictly disjoint and strictly ordered in time, so the test split measures genuine forward-in-time generalization to discharges run after every shot seen in training; it is the population used for all deployment-accuracy comparisons in this paper.

The RPNN processes its 53-dimensional input through the architecture illustrated in Fig.~\ref{fig:architecture-ablated}. Two output heads produce the predicted mean and log-variance of the next plasma state. A variance-pinning nonlinearity constrains the log-variance head to a fixed interval, preventing the predictive uncertainty from collapsing or diverging during training.

Table~\ref{tab:param-breakdown} summarizes the parameter count of each model
component for the full-size architecture. The residual-MLP decoder dominates the
parameter budget, accounting for nearly three-quarters of all trainable weights. This motivates the model reduction study later described in Sec.~\ref{sec:size-ablation}.

\begin{table}[ht]
    \caption{Trainable parameter count by component for the full-size RPNN architecture.}
    \label{tab:param-breakdown}
    \begin{ruledtabular}
    \begin{tabular}{lrc}
    \textrm{Component} & \textrm{Parameters} & \textrm{Share} \\
    \colrule
    Encoder MLP                    & $290{,}304$   & $8.4\%$ \\
    BatchNorm layer   & $2{,}048$     & $0.1\%$ \\
    GRU                            & $591{,}360$   & $17.1\%$ \\
    Decoder (residual MLP)         & $2{,}560{,}640$ & $74.2\%$ \\
    Mean / log-variance heads      & $6{,}966$     & $0.2\%$ \\
    Variance pinning                & $54$          & $\sim\!0.0\%$ \\
    \colrule
    Trainable total                & $3{,}451{,}372$ & $100\%$ \\
    \end{tabular}
    \end{ruledtabular}
\end{table}

We validate the LayerNorm$\to$BatchNorm substitution via two independent tests, both using our own retrained Keras checkpoints for each normalization type. Under an identical training curriculum and data split and compare them on the validation split (Table~\ref{tab:norm-train}): BatchNorm converges
to a better optimum ($-3.9\%$ MSE, $+4.0\%$ EV). We then evaluate both
checkpoints on the held-out test split (Table~\ref{tab:norm-eval}): BatchNorm again leads ($-1.3\%$ MSE, $+5.1\%$ EV). The direction of the effect is consistent across both tests, indicating that the substitution required for \textsc{hls4ml}/QKeras compatibility carries no accuracy cost at all. For the remainder of this work, we adopt BatchNorm as the baseline architecture choice.

\begin{table}[ht]
    \footnotesize
    \caption{Normalization-type training comparison: both variants trained from scratch
        under an identical curriculum and evaluated on the same chronological
        validation split ($982$ shots, $2{,}407$ sequences).}
    \label{tab:norm-train}
    \begin{ruledtabular}
    \begin{tabular}{lrr}
    \textrm{Norm type} & \textrm{MSE} & \textrm{EV} \\
    \colrule
    LayerNorm & $0.02728$ & $0.4967$ \\
    BatchNorm & $0.02622$ & $0.5166$ \\
    \colrule
    $\Delta$ (BatchNorm vs.\ LayerNorm) & $-3.9\%$ & $+4.0\%$ \\
    \end{tabular}
    \end{ruledtabular}
\end{table}

\begin{table}[ht]
    \footnotesize
    \caption{Normalization-type evaluation of  Keras checkpoints trained from scratch (same models as Table~\ref{tab:norm-train}),
        evaluated on the held-out test split.}
    \label{tab:norm-eval}
    \begin{ruledtabular}
    \begin{tabular}{lrr}
    \textrm{Norm type} & \textrm{MSE} & \textrm{EV} \\
    \colrule
    LayerNorm & $0.02407$ & $0.4038$ \\
    BatchNorm & $0.02375$ & $0.4243$ \\
    \colrule
    $\Delta$ (BatchNorm vs.\ LayerNorm) & $-1.3\%$ & $+5.1\%$ \\
    \end{tabular}
    \end{ruledtabular}
\end{table}

\subsection{Two-stage Training Curriculum}
\label{sec:training-curriculum}

The RPNN, in both its floating-point and quantized forms, is trained using a two-stage
curriculum. In the first stage, the full model is randomly initialized and trained end-to-end to minimize the mean-squared error (MSE) between the predicted and true next plasma
state; because this loss does not depend on the predicted log-variance, the log-variance
head and variance-pinning bounds are not meaningfully trained in this stage. In the
second stage, the stage-1 checkpoint is loaded and all parameters are frozen except for
the log-variance head and the variance-pinning bounds, which are fine-tuned using a
heteroscedastic negative-log-likelihood (NLL) loss augmented with a regularization term
(coefficient $10^{-3}$) that penalizes an excessively wide variance-pinning interval.
Both stages use the AdamW optimizer (learning rate $3\times10^{-4}$, weight decay
$10^{-3}$, batch size $512$) for up to $1{,}000$ epochs, with early stopping on the
validation loss (patience $250$ epochs). Since the model utilizes a gated recurrent unit (GRU) which relies on a hidden state that need to be initialized, the models are trained on full-length sequences
with each sequence's first two
timesteps excluded from the loss. In this way the initial uninformative GRU hidden state does not contribute towards the gradient.

Quantization-aware training extends this same two-stage curriculum to the quantized
model: a floating-point checkpoint is used to initialize the
QKeras~\cite{coelho2021qkeras,qkeras_software} model's weights, which are then fine-tuned
end to end under simulated fixed-point precision at a given target bit-width, following
the same staging as the floating-point training run.

An ensemble of RPNN members is trained by
repeating this curriculum with a distinct bootstrap resample of the training discharges
and a distinct random weight initialization per member. This provides the ensemble-based
estimate of epistemic uncertainty described in Sec.~\ref{sec:architecture}, in the spirit
of standard deep-ensemble uncertainty quantification~\cite{lakshminarayanan2017ensembles}.

\section{Ablation and Quantization}
\label{sec:ablation-quant}

\subsection{Architecture Size Ablation}
\label{sec:size-ablation}

Because the decoder accounts for a disproportionate share of the parameter budget (Table~\ref{tab:param-breakdown}), we systematically reduce the model by varying the decoder width (\texttt{decoder\_hidden\_dim}) and number of residual blocks (\texttt{decoder\_num\_res\_blocks}, default 3), followed by the GRU width (\texttt{gru\_hidden\_dim}, default 256) and the shared backbone width (\texttt{hidden\_dim}, default 512). 

Each configuration is trained from scratch using the same two-stage curriculum described in Sec.~\ref{sec:training-curriculum} and evaluated on two distinct populations, both drawn from the
 split detailed in Sec.~\ref{sec:architecture}. The first is the validation split
($982$ shots, $2{,}407$ sequences); because every configuration is trained and
validated under an identical curriculum, data split, and hyperparameters, this split
supports a training-matched comparison that isolates the effect of the architecture
change alone. The second is the held-out test split ($990$ most-recent shots,
$3{,}000$ sequences), never touched during training or model selection, used in a
second, independent evaluation pass to test whether conclusions drawn from the
training-matched comparison generalize forward in time; it is the population used for
all deployment-accuracy comparisons in this paper. Absolute metric values are not
directly comparable between the two splits.
Results below give each configuration's percent change in mean-squared error
(MSE) and explained variance (EV) relative to the full-size baseline evaluated on the
\emph{same} split, and the two evaluation passes are kept strictly separate: first the
training-matched comparison across all configurations
(Table~\ref{tab:ablation-pop1}), and then a second, independent
re-evaluation of the same configurations on the test split
(Table~\ref{tab:ablation-eval}).

Explained variance, used throughout this paper, is defined following
Ref.~\cite{char2024fullshot} as
\begin{equation}
    \mathrm{EV} := 1 - \frac{\mathrm{Var}(y - \hat{y})}{\mathrm{Var}(y)},
    \label{eq:ev}
\end{equation}
and computed for each of the 27 real-unit output channels before averaging
across channels, with $y$ the true target and $\hat{y}$ the model's mean prediction.
Unless otherwise noted, $y$ is the plasma state at the next timestep
used throughout Secs.~\ref{sec:algorithm}--\ref{sec:ablation-quant}. Sec.~\ref{sec:ev-rollout} reports a
rollout-horizon variant of the same metric, in which the model's accumulated
full-state prediction is scored against the true state at each horizon of a
free-running rollout.

\begin{table}[ht]
    \footnotesize
    \caption{Architecture size ablation, training-matched comparison on the
        chronological validation split, relative to
        the full-size baseline ($3{,}451{,}532$ params, MSE$=0.02622$, EV$=0.5166$).
        Naming: \texttt{dec}$N$ sets decoder\_hidden\_dim$=N$;
        \texttt{blocks1}/\texttt{b}$K$ set decoder\_num\_res\_blocks$=1$/$K$
        (default 3); \texttt{gru}$N$ sets gru\_hidden\_dim$=N$ (default 256);
        \texttt{hid}$N$ sets the shared backbone width hidden\_dim$=N$ (default 512).}
    \label{tab:ablation-pop1}
    \begin{ruledtabular}
    \begin{tabular}{lrrr}
    \textrm{Configuration} & $\Delta$\textrm{Params} & $\Delta$\textrm{MSE} & $\Delta$\textrm{EV} \\
    \colrule
    \texttt{dec384}                     & $-29.9\%$ & $-0.8\%$  & $-0.1\%$  \\
    \texttt{blocks1}                    & $-30.4\%$ & $+2.9\%$  & $-2.0\%$  \\
    \texttt{dec256}                     & $-52.3\%$ & $-2.2\%$  & $+3.0\%$  \\
    \texttt{dec256\_blocks1}            & $-59.9\%$ & $-0.8\%$  & $+2.2\%$  \\
    \texttt{gru128\_dec256}             & $-63.2\%$ & $+2.0\%$  & $-0.4\%$  \\
    \texttt{dec256\_b0}                 & $-63.7\%$ & $-1.1\%$  & $+1.7\%$  \\
    \texttt{hid256\_dec256}             & $-66.0\%$ & $-5.0\%$  & $+2.6\%$  \\
    \texttt{dec128\_b1}                 & $-68.9\%$ & $+3.6\%$  & $-1.9\%$  \\
    \texttt{gru128\_dec256\_b1}         & $-70.8\%$ & $+2.3\%$  & $+2.7\%$  \\
    \texttt{hid256\_dec256\_b1}         & $-73.6\%$ & $-1.1\%$  & $+0.8\%$  \\
    \texttt{hid384\_gru128\_dec256\_b1} & $-76.8\%$ & $+1.1\%$  & $+0.5\%$  \\
    \texttt{hid256\_gru128\_dec256\_b2} & $-77.9\%$ & $+1.7\%$  & $+1.1\%$  \\
    \texttt{gru128\_dec128\_b0}         & $-80.4\%$ & $+3.6\%$  & $-0.4\%$  \\
    \texttt{hid256\_gru128\_dec256\_b1} & $-81.7\%$ & $-0.0\%$  & $+1.2\%$  \\
    \texttt{gru64\_dec128\_b0}          & $-84.5\%$ & $+4.4\%$  & $-2.3\%$  \\
    \texttt{hid128\_gru128\_dec256\_b1} & $-85.7\%$ & $-1.8\%$  & $+0.8\%$  \\
    \texttt{gru64\_dec64\_b0}           & $-86.6\%$ & $+6.3\%$  & $-4.6\%$  \\
    \texttt{hid128\_gru128\_dec128\_b1} & $-92.9\%$ & $+1.5\%$  & $+1.0\%$  \\
    \texttt{hid128\_gru64\_dec128\_b3}  & $-93.0\%$ & $+4.9\%$  & $-5.4\%$  \\
    \texttt{hid128\_gru64\_dec128\_b2}  & $-94.0\%$ & $+4.8\%$  & $-2.9\%$  \\
    \texttt{hid128\_gru64\_dec128\_b1}  & $-94.9\%$ & $+4.6\%$  & $-2.7\%$  \\
    \texttt{hid64\_gru64\_dec64\_b1}    & $-97.9\%$ & $+12.4\%$ & $-10.5\%$ \\
    \texttt{hid64\_gru32\_dec64\_b1}    & $-98.4\%$ & $+22.9\%$ & $-17.9\%$ \\
    \texttt{hid32\_gru32\_dec32\_b1}    & $-99.2\%$ & $+31.0\%$ & $-29.5\%$ \\
    \texttt{hid32\_gru16\_dec32\_b1}    & $-99.3\%$ & $+35.5\%$ & $-32.1\%$ \\
    \end{tabular}
    \end{ruledtabular}
\end{table}

On the training-matched comparison, size reduction is remarkably cheap: every
configuration down to roughly $93\%$ parameter reduction stays within $\pm 5\%$ of the
full-size baseline's MSE, and no configuration
above the $73{,}228$-param mark loses more than $6.3\%$ MSE. This indicates meaningful
over-parameterization of the original architecture. Degradation only becomes clear and monotonic
once the shared backbone width is cut to 64 or 32 units: the four smallest
configurations (\texttt{hid64\_gru64\_dec64\_b1} through
\texttt{hid32\_gru16\_dec32\_b1}) show a clean, worsening trend from $+12.4\%$ MSE at
$73{,}228$ params to a real accuracy floor ($+35.5\%$ MSE at $22{,}508$ params).

\begin{table}[ht]
    \footnotesize
    \caption{The same architecture size ablation configurations as
        Table~\ref{tab:ablation-pop1}, evaluated on the held-out chronological test
        split ($990$ most-recent shots, $3{,}000$ sequences, never touched during
        training or model selection), relative to the full-size baseline on this split
        (MSE$=0.02375$, EV$=0.4243$). This is the population used for all
        deployment-accuracy comparisons in this paper. Rows are ordered by performance
        (best to worst $\Delta$MSE), not by parameter count as in
        Table~\ref{tab:ablation-pop1}.}
    \label{tab:ablation-eval}
    \begin{ruledtabular}
    \begin{tabular}{lrrr}
    \textrm{Configuration} & $\Delta$\textrm{Params} & $\Delta$\textrm{MSE} & $\Delta$\textrm{EV} \\
    \colrule
    \texttt{dec256}                     & $-52.3\%$ & $-1.5\%$  & $+4.2\%$  \\
    \texttt{gru128\_dec256}             & $-63.2\%$ & $-1.3\%$  & $+2.0\%$  \\
    \texttt{gru128\_dec256\_b1}         & $-70.8\%$ & $-0.7\%$  & $-0.1\%$  \\
    \texttt{hid256\_dec256}             & $-66.0\%$ & $-0.2\%$  & $+3.0\%$  \\
    \texttt{hid256\_dec256\_b1}         & $-73.6\%$ & $-0.2\%$  & $+1.6\%$  \\
    \texttt{hid128\_gru128\_dec256\_b1} & $-85.7\%$ & $+0.2\%$  & $-4.7\%$  \\
    \texttt{dec256\_blocks1}            & $-59.9\%$ & $+0.4\%$  & $+0.5\%$  \\
    \texttt{hid256\_gru128\_dec256\_b2} & $-77.9\%$ & $+0.9\%$  & $-1.8\%$  \\
    \texttt{hid384\_gru128\_dec256\_b1} & $-76.8\%$ & $+1.4\%$  & $-2.3\%$  \\
    \texttt{hid128\_gru64\_dec128\_b1}  & $-94.9\%$ & $+1.9\%$  & $-2.7\%$  \\
    \texttt{dec384}                     & $-29.9\%$ & $+2.9\%$  & $-0.3\%$  \\
    \texttt{hid256\_gru128\_dec256\_b1} & $-81.7\%$ & $+3.0\%$  & $-1.0\%$  \\
    \texttt{gru128\_dec128\_b0}         & $-80.4\%$ & $+4.2\%$  & $-5.0\%$  \\
    \texttt{hid128\_gru128\_dec128\_b1} & $-92.9\%$ & $+4.5\%$  & $-2.2\%$  \\
    \texttt{hid128\_gru64\_dec128\_b3}  & $-93.0\%$ & $+4.6\%$  & $-7.6\%$  \\
    \texttt{hid128\_gru64\_dec128\_b2}  & $-94.0\%$ & $+5.0\%$  & $-4.3\%$  \\
    \texttt{blocks1}                    & $-30.4\%$ & $+5.5\%$  & $-4.0\%$  \\
    \texttt{hid64\_gru64\_dec64\_b1}    & $-97.9\%$ & $+8.9\%$  & $-15.6\%$ \\
    \texttt{dec128\_b1}                 & $-68.9\%$ & $+15.3\%$ & $-14.9\%$ \\
    \texttt{hid64\_gru32\_dec64\_b1}    & $-98.4\%$ & $+19.9\%$ & $-26.2\%$ \\
    \texttt{hid32\_gru32\_dec32\_b1}    & $-99.2\%$ & $+22.5\%$ & $-37.6\%$ \\
    \texttt{hid32\_gru16\_dec32\_b1}    & $-99.3\%$ & $+24.8\%$ & $-40.6\%$ \\
    \texttt{dec256\_b0}                 & $-63.7\%$ & $+26.9\%$ & $-23.7\%$ \\
    \texttt{gru64\_dec128\_b0}          & $-84.5\%$ & $+54.3\%$ & $-44.9\%$ \\
    \texttt{gru64\_dec64\_b0}           & $-86.6\%$ & $+54.4\%$ & $-45.9\%$ \\
    \end{tabular}
    \end{ruledtabular}
\end{table}

Re-evaluating on the held-out test split shows removing \emph{all} residual blocks
is disproportionately damaging to forward-in-time generalization:
\texttt{dec256\_b0}, \texttt{gru64\_dec128\_b0}, and \texttt{gru64\_dec64\_b0} sit at
$+26.9\%$ to $+54.4\%$ MSE; keeping even a single residual block avoids this
collapse at every size. We also observe that decoder-width reduction remains the most
favorable knob; the \texttt{dec256}-family configurations, with $-73.6\%$ params while maintaining accuracy.
As on the training-matched split, the most
aggressive backbone-width reductions show a clear, worsening accuracy floor, running
from $+8.9\%$ MSE at $73{,}228$ params to $+24.8\%$ MSE at $22{,}508$ params.

Based on this trade-off, \texttt{hid128\_gru64\_dec128\_b1} ($175{,}628$ params,
$-94.9\%$) is the configuration selected for hardware deployment. Its halved GRU width
substantially reduces the FPGA resource cost identified in the \textsc{hls4ml} synthesis
study (Sec.~\ref{sec:firmware}) which the parameter count alone does not
capture. The accuracy price of the near-maximal parameter reduction is modest
on both splits: $+4.6\%$ MSE / $-2.7\%$ EV on the training-matched comparison and
$+1.9\%$ MSE / $-2.7\%$ EV on the held-out test split. Notably, 
increasing the width does not increase model accuracy. The \texttt{b2} and
\texttt{b3} variants of the same configuration are no better on either split despite
being $19$--$38\%$ larger. The selected configuration is deployed with 16-bit
quantization-aware training targeting an Alveo U50 device.

\begin{figure*}[t]
    \centering
    \resizebox{\textwidth}{!}{%
\begin{tikzpicture}[
    >={Latex[scale=0.75]}, thick,
    merge/.style={draw, rounded corners, fill=violet!30, minimum width=0.9375cm, minimum height=2.75cm},
    normC/.style={draw, rounded corners, fill=teal!30, minimum width=1.0cm, minimum height=1.0cm, font=\scriptsize, align=center},
    gruC/.style={draw=green!45!black, rounded corners, fill={rgb,255:red,205;green,227;blue,196}, minimum width=1.6cm, minimum height=1.1cm, font=\normalsize, align=center},
    decC/.style={draw, rounded corners, fill=black!8, minimum width=0.9375cm, minimum height=2.75cm},
    head/.style={draw, rounded corners, fill=blue!25, minimum width=1.0cm, minimum height=1.8cm},
    varpin/.style={draw, rounded corners, fill=blue!10, minimum width=1.4cm, minimum height=0.9cm, font=\small, align=center},
    sumnode/.style={draw, circle, minimum size=0.45cm, inner sep=0pt, font=\normalsize, fill=white},
    lbl/.style={font=\normalsize},
    tag/.style={font=\footnotesize, align=center},
]

\node[merge] (enc) at (0,0) {};
\coordinate (encTop) at ($(enc.north west)!0.18!(enc.south west)$);
\coordinate (encMid) at ($(enc.north west)!0.50!(enc.south west)$);
\coordinate (encBot) at ($(enc.north west)!0.82!(enc.south west)$);

\node[anchor=east, lbl] (lblStates)   at ($(encTop)+(-3.4/9,0)$) {States};
\node[anchor=east, lbl] (lblAct)      at ($(encMid)+(-3.4/9,0)$) {Actuators};
\node[anchor=east, lbl] (lblNextAct)  at ($(encBot)+(-3.4/9,0)$) {Actuator $\Delta$};
\draw[->] (lblStates.east)  -- (encTop);
\draw[->] (lblAct.east)     -- (encMid);
\draw[->] (lblNextAct.east) -- (encBot);
\node[tag, below=1mm of enc.south] {$512\to128$};

\node[decC] at (7.3,0)      (dec0) {};
\node[decC] at (8.7375,0)  (dec1) {};
\node[sumnode] at (9.6375,0) (s1) {$\oplus$};
\node[decC] at (10.675,0)  (dec2) {};

\draw[->] (enc.east) -- (dec0.west);
\draw[->] (dec0) -- (dec1);
\draw (dec1) -- (s1);
\draw[->] (s1) -- (dec2);

\coordinate (railL) at ($(dec1.north)+(0,0.6)$);
\coordinate (railR) at (s1.north |- railL);
\draw (dec1.north) -- (railL);
\draw (railL) -- (railR);
\draw[->] (railR) -- (s1.north);

\node[tag, below=2mm of dec1.south] {$512\to128$\\$3\to1$ residual blocks};

\node[normC] at (2.5,0.875) (norm) {BatchNorm};
\node[gruC]  at (4.8,0.825) (gru)  {GRU};

\draw[->] (enc.east) -- (norm.west);
\draw[->] (norm) -- (gru);
\draw[->] (gru.east) -- (dec0.west);
\node[tag, above=2mm of gru.north] {$256\to64$};

\node[head] at (12.775,1.6)  (mean)   {};
\node[head] at (12.775,-1.6) (logvar) {};
\node[varpin] at (14.775,-1.6) (varpin) {Var.\\Pinning};

\draw[->] (dec2.east) -- (mean.west);
\draw[->] (dec2.east) -- (logvar.west);
\draw[->] (logvar) -- (varpin);
\node[anchor=west, lbl] (lblMean)   at ($(mean.east)+(0.8,0)$)   {$\mu$};
\node[anchor=west, lbl] (lblLogvar) at ($(varpin.east)+(0.8,0)$) {$\log(\sigma^2)$};
\draw[->] (mean.east) -- (lblMean.west);
\draw[->] (varpin.east) -- (lblLogvar.west);

\end{tikzpicture}%
}
    \caption{RPNN architecture, drawn with the layer widths of the deployed
        configuration (\texttt{hid128\_gru64\_dec128\_b1},
        Sec.~\ref{sec:size-ablation}); the full-size baseline shares the identical
        topology with wider layers (backbone $512$, GRU $256$, decoder $512$ with
        three residual blocks), and each block's width reduction relative to it is
        noted below the block (e.g.\ $512\!\to\!128$). States (27) and actuators (13,
        supplied both as their current absolute level and as the commanded delta into
        the following timestep, for a 53-dimensional encoder input) are normalized
        and encoded, then passed through a BatchNorm layer before a single-layer GRU;
        the GRU output is concatenated with the encoder's own output (bypassing the
        GRU) and passed to a residual-MLP decoder, whose output feeds two linear
        heads producing the predicted mean and log-variance, the latter passing
        through a variance-pinning nonlinearity.}
    \label{fig:architecture-ablated}
\end{figure*}
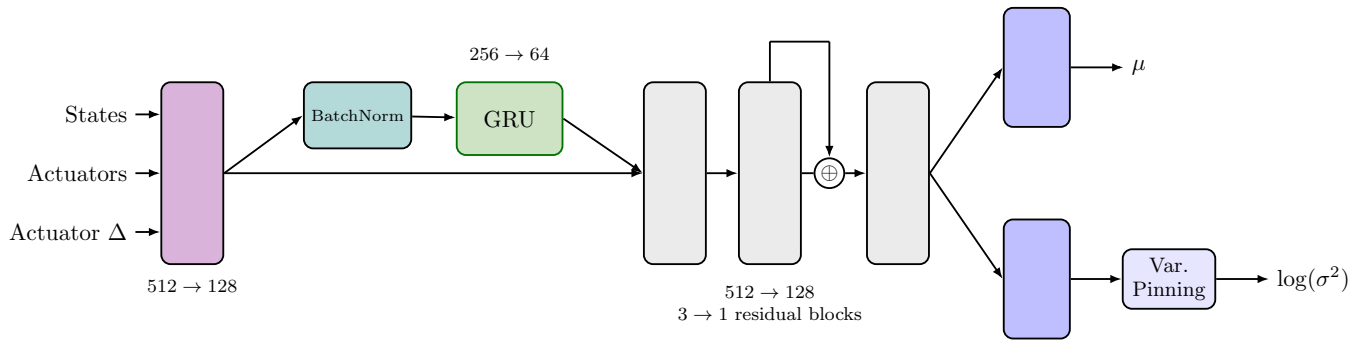

The deployed architecture (\texttt{hidden\_dim}$=128$, \texttt{gru\_hidden\_dim}$=64$,
\texttt{decoder\_hidden\_dim}$=128$, \texttt{decoder\_num\_res\_blocks}$=1$), shown in
Fig.~\ref{fig:architecture-ablated}, redistributes the parameter budget substantially
relative to the full-size model of Table~\ref{tab:param-breakdown}, as shown in
Table~\ref{tab:chosen-param-breakdown}.

\begin{table}[ht]
    \caption{Parameter count by pipeline component for the deployed architecture,
        \texttt{hid128\_gru64\_dec128\_b1} ($175{,}628$ total params, $-94.9\%$ relative
        to the full-size architecture of Table~\ref{tab:param-breakdown}), using the same
        convention as Table~\ref{tab:param-breakdown}.}
    \label{tab:chosen-param-breakdown}
    \begin{ruledtabular}
    \begin{tabular}{lrc}
    \textrm{Component} & \textrm{Parameters} & \textrm{Share} \\
    \colrule
    Encoder MLP                    & $23{,}424$  & $13.3\%$ \\
    BatchNorm layer   & $512$       & $0.3\%$ \\
    GRU                            & $37{,}248$  & $21.2\%$ \\
    Decoder (residual MLP)         & $107{,}264$ & $61.1\%$ \\
    Mean / log-variance heads      & $6{,}966$   & $4.0\%$ \\
    Variance pinning                & $54$        & $\sim\!0.0\%$ \\
    \colrule
    Trainable total                & $175{,}468$ & $100\%$ \\
    \end{tabular}
    \end{ruledtabular}
\end{table}

\subsection{Quantization-aware Training}
\label{sec:qat-sweep}

Having selected \texttt{hid128\_gru64\_dec128\_b1} as
the deployment architecture, we then quantize it via QKeras to the fixed-point
precision \texttt{ap\_fixed<16,6>}, using identical precision for weights,
activations, and biases, applying the same two-stage curriculum of
Sec.~\ref{sec:training-curriculum} under simulated fixed-point precision. This precision was selected in
preliminary sweeps, in which the more aggressive \texttt{ap\_fixed<8,3>} and
\texttt{ap\_fixed<6,1>} variants lost substantially more accuracy after
fine-tuning. Final accuracies are evaluated on the held-out test split, with the FP32 (unquantized) reference for the deployment architecture being
 MSE$=0.02420$, EV$=0.4128$.

Naive post-training quantization --- transplanting the FP32 weights into the 16-bit
architecture with no fine-tuning --- destroys the model outright (validation MSE
$1.469$, EV $-13.57$), while the two-stage fine-tuning recovers the accuracy,
indicating that quantization-aware training is a critical step in quantization and
FPGA deployment.

\begin{table}[ht]
    \footnotesize
    \caption{16-bit QAT on the chosen deployment architecture, final accuracy after
        fine-tuning, relative to the FP32 reference for this architecture
        (MSE$=0.02420$, EV$=0.4128$) on the held-out test split.}
    \label{tab:qat-eval}
    \begin{ruledtabular}
    \begin{tabular}{lrrrr}
    \textrm{Precision} & \textrm{MSE} & $\Delta$\textrm{MSE} & \textrm{EV} & $\Delta$\textrm{EV} \\
    \colrule
    \texttt{ap\_fixed<16,6>} & $0.02487$ & $+2.8\%$  & $0.3932$ & $-4.8\%$  \\
    \end{tabular}
    \end{ruledtabular}
\end{table}

After fine-tuning, the 16-bit quantized model lands within $+2.8\%$ MSE / $-4.8\%$ EV
of the FP32 reference on the held-out test split
(Table~\ref{tab:ablation-eval}). This is the configuration carried into hardware
synthesis (Sec.~\ref{sec:firmware}). Relative to the full-size FP32 baseline, the
combined cost of the $94.9\%$ architecture reduction and 16-bit quantization is
$+4.7\%$ MSE / $-7.3\%$ EV on the test split.

\section{Performance Measurements}
\label{sec:performance}

\subsection{Ensemble Replay of a Held-out shot}
\label{sec:rollout}

The population-level MSE/EV metrics of Sec.~\ref{sec:ablation-quant} are computed one step at a time, with the true previous state fed into the model at each step. Real deployment, however, is autoregressive: after the
first step, the model must condition on its own previous prediction, so errors can
compound over a full discharge. Following the ``replay'' evaluation of
Ref.~\cite{char2024fullshot}, we predict an entire held-out shot forward from only its
first true state: at every step the model receives the true actuator trajectory, but
the state input is entirely self-generated, with each rollout sampling its next state
from the predicted Gaussian at every step. As in the reference implementation's replay
path, each predicted state increment is clamped per channel to bounds derived from the
training split ($0.5/99.5$ percentiles of each channel's observed values and one-step
increments). In practice, this clamping engages rarely ($1.3\%$ of (step, channel) updates for the
baseline ensemble, $2.0\%$ for the quantized one).

We compare two ensembles rolled out this way: the full-size FP32 baseline (with BatchNorm,
Sec.~\ref{sec:architecture}) and the deployed \texttt{hid128\_gru64\_dec128\_b1}
architecture under 16-bit quantization-aware training.
Each ensemble comprises its full set of fifteen members, each trained on its own
bootstrap resample of the training shots with random initialization
(Sec.~\ref{sec:training-curriculum}). No stability screening was applied. For
each configuration we draw $45$ rollout samples, three from each member, so that
both the ensemble's epistemic spread and each member's own aleatoric sampling
contribute to the visualized uncertainty band.

\begin{figure*}[t]
    \centering
    \includegraphics[width=\textwidth,height=0.92\textheight,keepaspectratio]{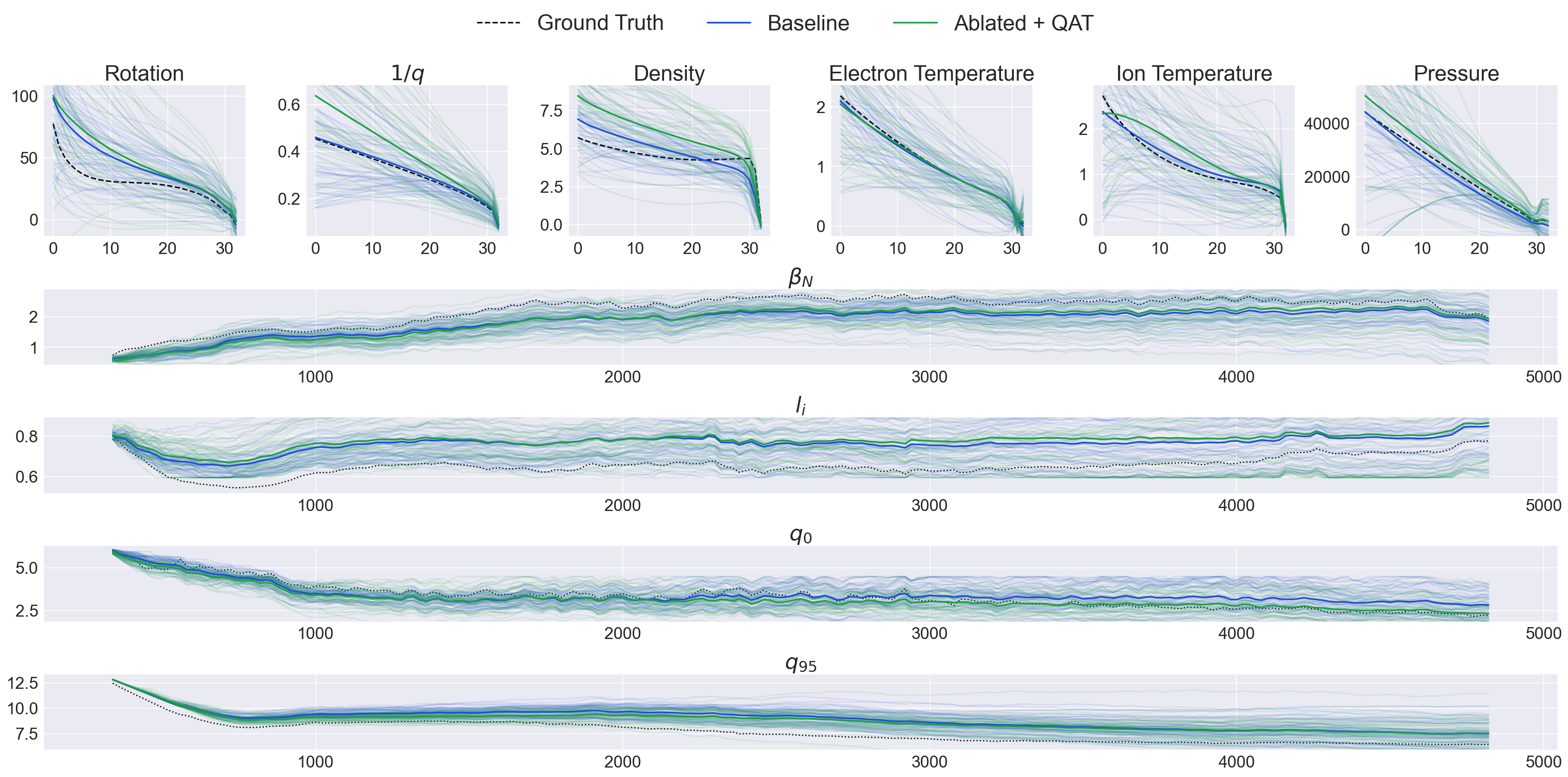}
    \caption{Fully autoregressive ensemble replay of held-out test shot $193369$
        ($225$ steps), FP32 baseline (blue, all 15 bootstrap members, 45 rollouts)
        vs.\ the deployed 16-bit \texttt{hid128\_gru64\_dec128\_b1} model (green,
        same convention), against ground truth (dashed black). Top row: all six
        profile quantities at the final timestep in physical units, vs.\ minor
        radius index (the safety-factor profile is stored, and shown, as $1/q$).
        Remaining rows: four representative scalar quantities ($\beta_N$, $l_i$,
        $q_0$, $q_{95}$) over the full shot, plotted against absolute time into the
        discharge ($320$--$4{,}820\,$ms, so the ramp-up is included); RMSE for these
        and the three not plotted here (\texttt{dssdenest}, \texttt{vloop},
        \texttt{wmhd\_EFIT01}) is reported in full in
        Table~\ref{tab:rollout-scalar}. Faint lines are individual rollout samples;
        bold lines are the ensemble mean. Vertical axes are clipped to the range of
        the ground truth and ensemble means, so a few individual sampled
        trajectories extend beyond the plotted range.}
    \label{fig:rollout}
\end{figure*}

The chosen shot has a large actuator driven range and illustrates model's capability to track trends. 
Figure~\ref{fig:rollout} shows the resulting replay: four representative scalar
quantities over the full shot duration, and all six profile quantities reconstructed
from their PCA coefficients at the final timestep. Table~\ref{tab:rollout-scalar} and Table~\ref{tab:rollout-profile} report
the corresponding root-mean-square error (RMSE, normalized units) of each ensemble's
mean trajectory against ground truth and at the final timestep for the six reconstructed
profiles.

\begin{table}[ht]
    \footnotesize
    \caption{Scalar-quantity RMSE (normalized units) over the full autoregressive
        rollout of shot $193369$ (16-bit vs.\ floating-point 32-bit precision).}
    \label{tab:rollout-scalar}
    \begin{ruledtabular}
    \begin{tabular}{lrrr}
    \textrm{Quantity} & \textrm{FP32} & \textrm{16-bit} & $\Delta$ \\
    \colrule
    \texttt{betan\_EFIT01} & $0.4068$ & $0.3775$ & $-7.2\%$  \\
    \texttt{dssdenest}     & $0.3010$ & $0.3125$ & $+3.8\%$  \\
    \texttt{li\_EFIT01}    & $0.5686$ & $0.6422$ & $+12.9\%$ \\
    \texttt{q0\_EFIT01}    & $0.5772$ & $0.4221$ & $-26.9\%$ \\
    \texttt{q95\_EFIT01}   & $0.6557$ & $0.5706$ & $-13.0\%$ \\
    \texttt{vloop}         & $0.4857$ & $0.5031$ & $+3.6\%$  \\
    \texttt{wmhd\_EFIT01}  & $0.2190$ & $0.1367$ & $-37.6\%$ \\
    \end{tabular}
    \end{ruledtabular}
\end{table}

\begin{table}[ht]
    \footnotesize
    \caption{Profile RMSE (normalized units, 33-point radial reconstruction) at the
        final timestep of the same rollout (ensemble mean vs.\ ground truth).}
    \label{tab:rollout-profile}
    \begin{ruledtabular}
    \begin{tabular}{lrrr}
    \textrm{Quantity} & \textrm{FP32} & \textrm{16-bit} & $\Delta$ \\
    \colrule
    \texttt{temp}         & $0.0864$ & $0.0593$ & $-31.4\%$  \\
    \texttt{itemp}        & $0.1664$ & $0.3034$ & $+82.3\%$  \\
    \texttt{dens}         & $0.5862$ & $0.6106$ & $+4.2\%$   \\
    \texttt{rotation}     & $0.3173$ & $0.4214$ & $+32.8\%$  \\
    \texttt{pres\_EFIT01} & $0.0788$ & $0.0287$ & $-63.6\%$  \\
    \texttt{q\_EFIT01}    & $0.0203$ & $0.0805$ & $+296.7\%$ \\
    \end{tabular}
    \end{ruledtabular}
\end{table}

On this shot, both ensembles track the large, actuator-driven trends of the discharge
across the full $225$-step, $4.5$-second replay (Fig.~\ref{fig:rollout}). 

\subsection{Explained Variance over Rollout Horizon}
\label{sec:ev-rollout}

To investigate how accuracy degrades as a function of rollout
horizon, we follow the rollout-horizon EV analysis in Fig.~3 of
Ref.~\cite{char2024fullshot}. We compute EV (Eq.~\ref{eq:ev}) as an explicit function
of the number of autoregressive steps taken since a rollout began, pooled across
starting points within every test shot. EV at horizon $t$ scores the model's
accumulated full-state prediction against the true state, $y = s_{s+t}$ and
$\hat{y} = \hat{s}_{s+t}$, with $s$ the rollout's starting timestep. Because the plasma
state moves far less over a short horizon than it varies across the population, a
persistence model that simply repeats $s_s$ at every timestep scores $0.96$ at $t=1$,
$0.55$ at $t=25$, and crosses zero only at $t \approx 99$. Model skill is therefore measured as the improvement over this baseline, rather than by the EV value. 

\begin{figure*}[!t]
    \centering
    \includegraphics[width=\textwidth]{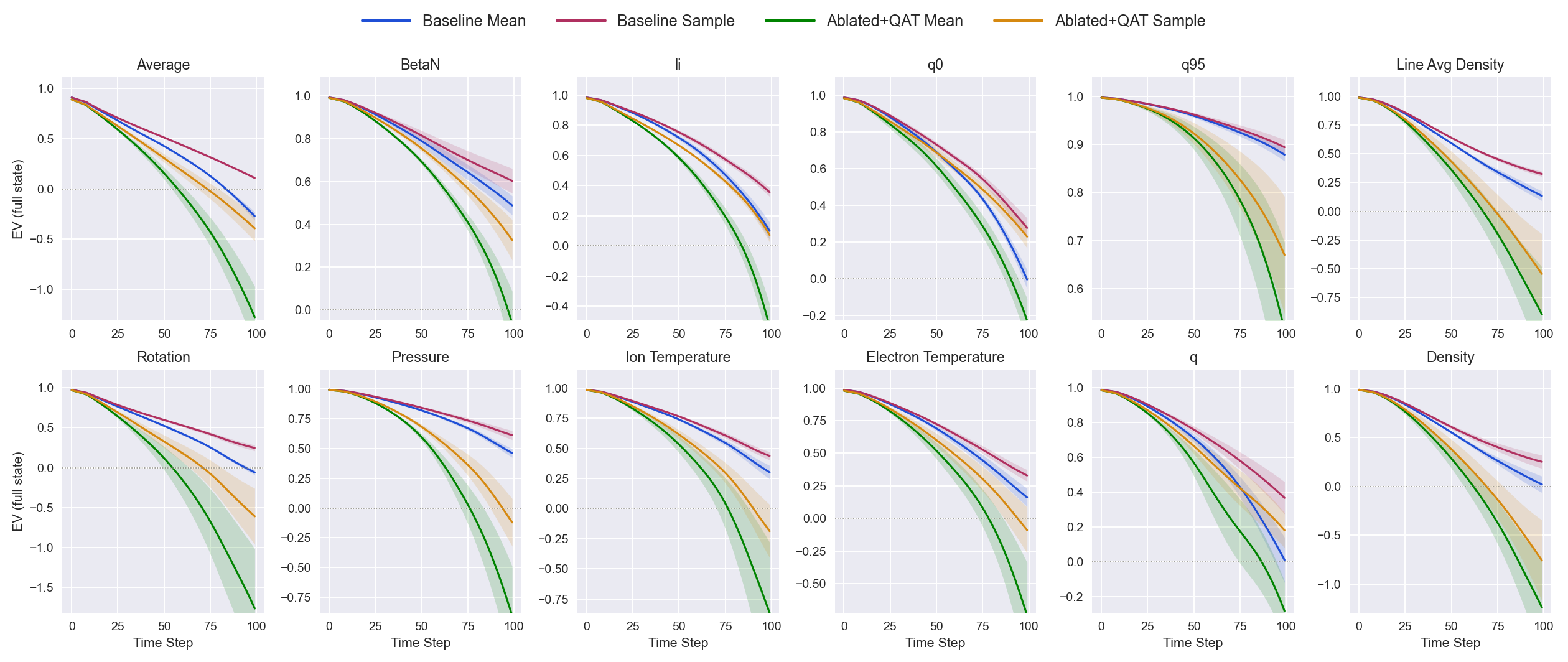}
    \caption{Full-state explained variance (Eq.~\ref{eq:ev} with $y = s_{s+t}$ and
        $\hat{y}$ the model's accumulated prediction) as a function of
        autoregressive rollout horizon over steps $0$--$100$, pooled over every
        starting position in every held-out test shot ($112{,}436$ (shot, start)
        combinations, $30$ sampled continuations each). Top row: the average across
        all $27$ output channels and five representative scalars ($\beta_N$, $l_i$,
        $q_0$, $q_{95}$, line-averaged density); bottom row: the first PCA component
        of all six profile quantities. Curves compare the full-size FP32 baseline
        (blue: mean mode; crimson: sample mode) against the deployed 16-bit
        \texttt{hid128\_gru64\_dec128\_b1} model (green: mean; orange: sample), each
        averaged over four members from the ensemble, with shaded bands showing the standard error
        across members. Curves are smoothed for
        display with a centered $15$-step rolling average; the underlying per-step
        values are unsmoothed.}
    \label{fig:ev-rollout}
\end{figure*}

For every test shot, we first make a single teacher-forced pass through its true
trajectory, caching the GRU hidden state entering each timestep; a rollout starting at
timestep $s$ thus begins from the hidden state a teacher-forced pass would
have built from the true steps $0,\dots,s-1$. From every start position within
each shot, the model is rolled out autoregressively for the remaining steps in that shot, supplied only with the true actuator trajectory and no state clamping
applied.
We compare two prediction modes at every step: ``Mean'' feeds back the model's predicted
mean deterministically, while ``Sample'' draws $30$ independent stochastic
continuations per combination from the predicted Gaussians, averaging the
accumulated trajectories before scoring.

Training uses only one-step teacher forcing and therefore does not penalize error growth during closed-loop rollouts. As a result, free-rollout stability varies substantially across ensemble members. Even with identical model configurations, differences in random initialization and bootstrap resampling can cause some members to diverge to negative EV at long horizons.

We therefore screen each member in both prediction modes. Figure~\ref{fig:rollout} reports results for these ensembles, while Fig.~\ref{fig:ev-rollout} shows their member-averaged EV over the first $100$ rollout steps, corresponding to $2,$s of plasma evolution.

The key takeaway of Fig.~\ref{fig:ev-rollout} is that the size-plus-quantization cost
compounds in closed loop: the quantized ensemble's curves fall away from the
baseline's at every horizon, and by $t=100$ they trail by $0.51$ EV in sample mode
($-0.41$ vs.\ $+0.10$). A one-step accuracy gap of under five percent
(Sec.~\ref{sec:qat-sweep}) thus grows into half an EV point within $100$
autoregressive steps ($2\,$s of plasma evolution) --- a closed-loop amplification
that single-step metrics cannot reveal, and a direct target for future
closed-loop-aware training. Sample mode also degrades more slowly than mean mode for
both ensembles, as averaging $30$ noisy continuations suppresses individually
diverging trajectories.

\subsection{Uncertainty Quantification}
\label{sec:uq}

The RPNN quantifies uncertainty through two complementary mechanisms
(Sec.~\ref{sec:architecture}): the per-channel predictive Gaussian captures aleatoric
uncertainty, and the ensemble of independently trained members captures epistemic
uncertainty. For the predictive distribution to be usable by a downstream controller,
these two sources must be combined into a single distribution over rollout
trajectories. We compare the three
combination methods ``Mean-TS1", ``Sample-TS1" and ``Sample-TSInf" used in Ref.~\cite{char2024fullshot} (first presented by 
Chua~et~al in~\cite{chua2018pets}). In ``Mean-TS1'', each rollout step feeds back the mean
of the predicted Gaussian, but the ensemble member generating the step is resampled
every timestep. In ``Sample-TS1'', each step instead samples from the predicted
Gaussian, again resampling the member every timestep. In ``Sample-TSInf'', each step
samples from the predicted Gaussian, but a single member, drawn once, generates the
entire trajectory. In every method all members propagate their own hidden states along
the realized trajectory; the method only governs whose output is emitted. Because the
autoregressive rollout does not have a simple analytical probability distribution, the predictive
distribution at each horizon is estimated for each channel from 30 sampled trajectories for each shot and starting point. For each predicted variable, the mean and standard deviation of these trajectories are used as an approximation of a gaussian distribution. This strategy is also used in the reference paper.

We evaluate the predictive distributions with the same two metrics as
Ref.~\cite{char2024fullshot}: the coverage of a $90\%$ prediction interval
(PI), and the miscalibration area. Given per-observation prediction intervals
$\mathrm{PI}_{n,(1-\alpha)}$ constructed to capture probability mass $1-\alpha$,
coverage is the empirical frequency of observations falling inside their intervals,
\begin{equation}
    \mathrm{Coverage}_{(1-\alpha)}
    = \frac{1}{N}\sum_{n=1}^{N} \mathbb{1}\{\,y_n \in \mathrm{PI}_{n,(1-\alpha)}\,\},
    \label{eq:coverage}
\end{equation}
ideally equal to $1-\alpha$ itself. Our reported prediction intervals are the
empirical $5\%$/$95\%$ quantile band of the $30$ sampled trajectories, exactly as in
the reference implementation. Coverage probes
a single interval; the miscalibration area aggregates the same deviation over all of
them: for a set of $M$ expected probabilities $p_i$ spanning $[0,1]$,
\begin{equation}
    \mathrm{MA} = \frac{1}{M}\sum_{i \in [M]}
    \bigl|\, p_i - \mathrm{Coverage}_{p_i} \,\bigr|,
    \label{eq:miscal}
\end{equation}
so that $\mathrm{MA}=0$ is perfect calibration at every confidence level, higher is
worse, and $\approx 0.5$ is the theoretical maximum. We compute it with the
Uncertainty Toolbox~\cite{chung2021uncertainty}, as in the reference. Both metrics,
like EV, are computed per output channel and per rollout horizon on the identical
population, protocol, and rollouts as Sec.~\ref{sec:ev-rollout}, with
horizons extended to the data ceiling $t=277$; metrics are then averaged across the
$27$ channels. The EV panel
uses the reference's change-from-start convention, whose no-skill floor is zero at
every horizon (Sec.~\ref{sec:ev-rollout}).

\begin{figure*}[!t]
    \centering
    \includegraphics[width=\textwidth]{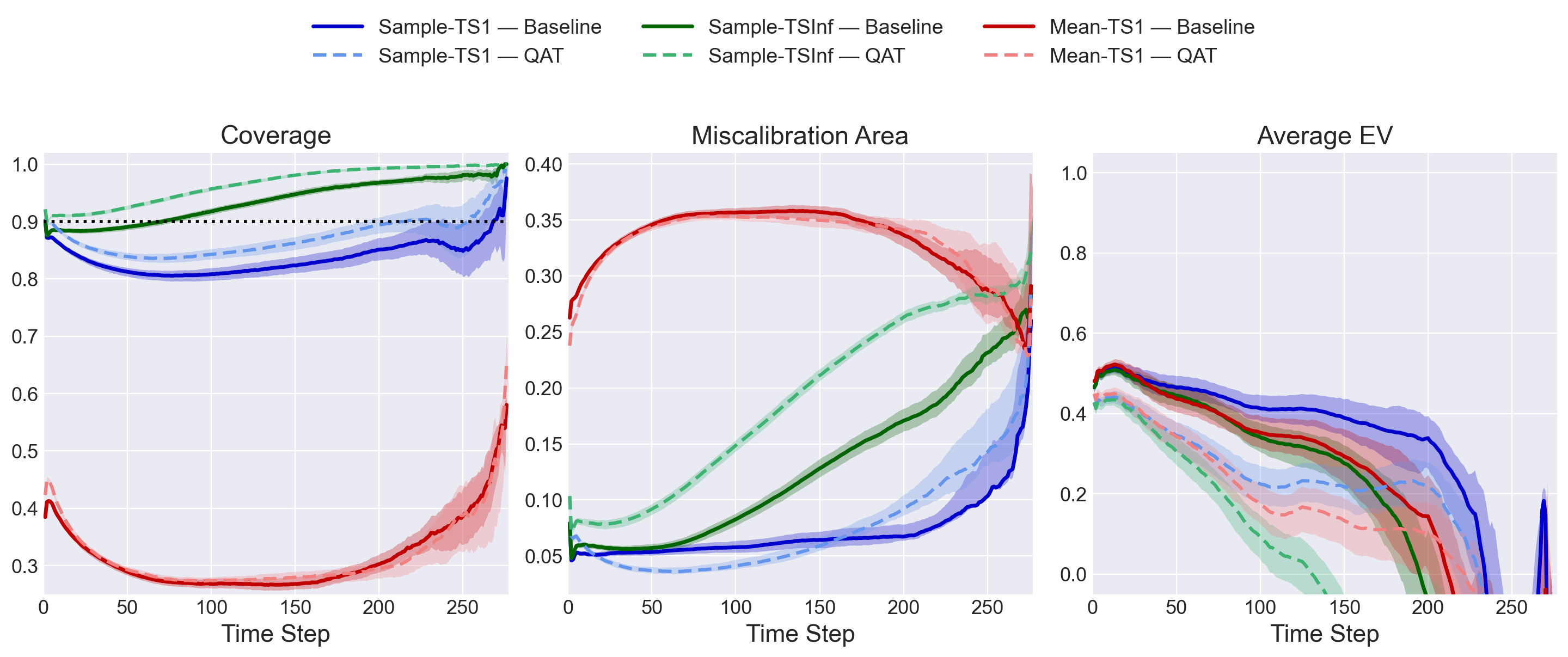}
    \caption{Uncertainty metrics over rollout horizon for the three
        predictive-distribution methods (blue: Sample-TS1; green: Sample-TSInf;
        red: Mean-TS1, the reference paper's palette), comparing the full-size FP32
        baseline ensemble (dark, solid; members $\{1,2,3,7\}$) against the deployed
        16-bit \texttt{hid128\_gru64\_dec128\_b1} ensemble (light, dashed; members
        $\{1,4,7,12\}$) within each method's hue. Panels, left to right: empirical
        coverage of the $90\%$ prediction interval (Eq.~\ref{eq:coverage}, dotted
        line marks the $0.90$ target); miscalibration area (Eq.~\ref{eq:miscal},
        lower is better); EV averaged over all $27$ output channels (axis clipped
        below $0$). Shaded bands are $90\%$ shot-level cluster-bootstrap confidence
        intervals ($B=200$) of each mean curve, computed from identical resamples
        for every curve so visible gaps are not resampling noise. The horizon
        extends to the data ceiling $t=277$, but the active population collapses
        past the $200$-step mark of Fig.~\ref{fig:ev-rollout} --- $n=3{,}508$
        combinations at $t=200$, $1{,}213$ at $t=225$, $283$ at $t=250$, $7$ at
        $t=275$ --- so only the longest shots remain and the tail beyond
        $t\approx250$ is qualitative (the brief EV spike near $t\approx267$ is the
        $n=7$ population, not a recovery).}
    \label{fig:uq}
\end{figure*}

Figure~\ref{fig:uq} reports coverage, miscalibration area, and average EV for the
three methods and both ensembles in a single figure. Each ensemble uses four members
 drawn from those passing the rollout-stability
screen of Sec.~\ref{sec:ev-rollout}. Since only one screened ensemble exists per configuration,
the standard error of the reference across models is not available; the shaded bands are
instead $90\%$ cluster-bootstrap confidence intervals over the $885$ test shots
contributing combinations, quantifying test-population sampling uncertainty of each curve.

The separation among the three methods match that reported in Ref.~\cite{char2024fullshot}. Mean-TS1 produces an overconfident prediction because it ignores the aleatoric noise, leading to coverage falling from $\approx 0.4$ at $t=1$ to $\approx 0.27$--$0.30$
through the second half of the horizon (against the $0.90$ target). Its
miscalibration area climbs to $\approx 0.35$, and its EV suffers as well (baseline
$0.14$ vs.\ $0.34$ for Sample-TS1 at $t=200$). Sample-TS1 is well calibrated over the
entire trusted horizon. Coverage stays within $0.81$--$0.90$ (baseline) and
$0.84$--$0.92$ (quantized) of the target, and miscalibration area at or below $0.08$
and $0.10$ respectively through $t=200$ while also achieving the best EV at
every horizon. Sample-TSInf matches Sample-TS1 at short horizons, but drifts into
increasing over coverage ($\to 0.97$--$0.99$ by $t=200$) with miscalibration
area growing to $0.17$ (baseline) and $0.26$ (quantized), and its EV falls to or
below zero by $t=200$ (baseline $-0.06$, quantized $-0.79$). We observe that a single fixed member's
sampled trajectory can wander far from the truth, and the $30$-sample empirical mean
is not robust to such excursions. This failure mode is suppressed by Sample-TS1's
per-step member resampling. Sample-TS1 therefore dominates on every
metric, and we adopt it as the recommended
inference mode for deployment. 
\section{Firmware Implementation}
\label{sec:firmware}

\subsection{\textsc{HLS4ML} Synthesis Workflow}
\label{sec:hls-workflow}

We synthesize the deployed \texttt{hid128\_gru64\_dec128\_b1} model (16-bit QAT,
Sec.~\ref{sec:qat-sweep}) via \textsc{hls4ml}~\cite{Duarte:2018ite,khoda2022ultralow},
translating the QKeras model into HLS C++ and carrying it through the full Vitis
HLS~$\to$~Vivado~$\to$~v++ implementation flow (Vitis/Vivado 2022.1), targeting a Xilinx
Alveo U50 accelerator card (part \texttt{xcu50-fsvh2104-2-e}, platform
\texttt{xilinx\_u50\_gen3x16\_xdma\_5\_202210\_1}).  The normalizer and variance-pinning nonlinearity
(Sec.~\ref{sec:architecture}) sit outside the synthesized dataflow region; the HLS
subgraph itself covers $175{,}414$ of the model's $175{,}628$ parameters.

We make the recurrent state explicit
kernel I/O: the hidden state enters the model as a second input alongside the
$53$-dimensional state/actuator vector, and the updated state is returned alongside
the mean and log-variance heads, so a single inference streams $117$ input words
($x$: $53$, $h_{\mathrm{prev}}$: $64$) and $118$ output words (mean: $27$,
log-variance: $27$, $h_{\mathrm{next}}$: $64$), where a ``word'' is one $32$-bit
transfer on the kernel's streaming interface, carrying one input or output value
per clock cycle. The host threads the state between
invocations and supplies zeros at shot start, giving exact episode reset with no
persistent state inside the FPGA. 

We found that the reuse factor (RF), which specifies how many times each multiply-accumulate unit is
time-multiplexed across the design's arithmetic operations, is the dominant lever
trading resource utilization against latency. We adopt $\mathrm{RF}=64$ with hls4ml's
resource strategy, streaming I/O, and a widened accumulator
precision (\texttt{ap\_fixed<32,20>}, versus the model's native 16-bit weights and
activations) to preserve numerical fidelity through the GRU/decoder chain. Preserving
that fidelity on this dataset additionally required three range-aware precision
corrections: a $27$-bit input type
(\texttt{ap\_fixed<27,17>}) as well as
wide types on the identity passthrough nodes \textsc{hls4ml} inserts after dense
layers. With these items in place, we simulate the complete kernel in C. The resulting model's accuracy reproduces the trained QKeras model's accuracy.

\subsection{Resource and Latency Trade-offs}
\label{sec:firmware-results}

Table~\ref{tab:firmware-resources} reports the resource utilization of the
implemented FPGA kernel after complete placement and routing (``post-route'', i.e.\
measured on the final implemented design rather than estimated at the synthesis
stage), relative to the user-accessible resource budget. All four resource categories fit comfortably within budget, with DSP the tightest constraint at $65.0\%$.

\begin{table}[ht]
    \footnotesize
    \caption{Post-route resource utilization of the deployed stateful
        \texttt{hid128\_gru64\_dec128\_b1} kernel (16-bit QAT, $\mathrm{RF}=64$, GRU
        hidden state as explicit kernel I/O) on the Alveo U50, relative to the
        user-accessible resource budget.}
    \label{tab:firmware-resources}
    \begin{ruledtabular}
    \begin{tabular}{lrrr}
    \textrm{Resource} & \textrm{Used} & \textrm{Budget} & \textrm{Util.} \\
    \colrule
    LUT      & $262{,}381$ & $744{,}638$     & $35.24\%$ \\
    FF (REG) & $280{,}955$ & $1{,}579{,}376$ & $17.79\%$ \\
    DSP      & $3{,}868$   & $5{,}948$       & $65.03\%$ \\
    BRAM     & $562$       & $1{,}163$       & $48.32\%$ \\
    URAM     & $0$         & $636$           & $0.0\%$   \\
    \end{tabular}
    \end{ruledtabular}
\end{table}

We report an achieved clock frequency of
$104.7\,$MHz, or $9.55\,$ns/cycle for the synthesiszed FPGA kernel.
Table~\ref{tab:firmware-latency} reports the design's per-component latency.
Purely structural operations (activations, concatenations, the residual add) fuse
into the stream pipeline and add zero incremental latency, and the rows do not sum
to the totals: the dataflow region overlaps adjacent stages, so each row is that
stage's own processing depth, while the Total and Interval rows are independently
tool-reported for the top level. End to end, one inference takes $276$ cycles
($2.64\,\mu$s), with an initiation interval of $122$ cycles ($1.17\,\mu$s,
$\sim\!8.6\times10^5$ inferences per second) between successive invocations. As no physical Alveo U50 card was available to us for direct hardware testing, this achieved-frequency figure is obtained directly from the post-implementation timing simulation by Vivado. 

To place these numbers in context, Table~\ref{tab:latency-benchmark} benchmarks the
same single-timestep inference in software: the FP32 TensorFlow implementations of
the full-size baseline and the deployed architecture, on an 8-core AMD
EPYC~7763 allocation and on a dedicated NVIDIA A100 80\,GB GPU, each compiled as a \texttt{tf.function}
and timed over $1{,}000$ consecutive invocations after warm-up.  On the
single-inference metric, the deployed FPGA kernel's $2.64\,\mu$s is
$\approx 240\times$ faster than the same architecture on the CPU and
$\approx 410\times$ faster than on the dedicated A100. We note that, while the per-sample latency for batched CPU/GPU execution approaches the FPGA implementation's frequency, the overall inference time remains prohibitive for online MPC controllers. 
\\
\\
\begin{table}[ht!]
    \footnotesize
    \caption{Software-vs-FPGA inference cost for one control-loop tick. Software
        rows: median wall-clock per call divided by batch size (in $\mu$s per
        sample), for the FP32 TensorFlow implementations of the full-size baseline
        ($3.45$\,M parameters) and the deployed ablated architecture ($176$\,K
        parameters), on an 8-core AMD EPYC~7763 allocation and a dedicated NVIDIA
        A100 80\,GB (PCIe, exclusive use), timed as a compiled \texttt{tf.function}
        with device synchronization included. Batch size $1$ is the operative
        regime for closed-loop control, where each tick depends on the previous
        prediction and cannot be batched. FPGA rows: the deployed 16-bit stateful
        kernel of Table~\ref{tab:firmware-latency}.}
    \label{tab:latency-benchmark}
    \begin{ruledtabular}
    \begin{tabular}{lrrrr}
    & \multicolumn{2}{c}{\textrm{Baseline (3.45M)}} & \multicolumn{2}{c}{\textrm{Ablated (176K)}} \\
    \textrm{Batch} & \textrm{CPU} & \textrm{A100} & \textrm{CPU} & \textrm{A100} \\
    \colrule
    $1$      & $1{,}011$ & $1{,}249$ & $608$  & $1{,}085$ \\
    $16$     & $106$     & $81.0$    & $45.0$ & $67.3$    \\
    $128$    & $34.2$    & $10.1$    & $9.9$  & $8.7$     \\
    $1{,}024$ & $24.9$   & $1.45$    & $2.83$ & $1.08$    \\
    \colrule
    FPGA, end-to-end latency    & \multicolumn{4}{c}{$2.64\,\mu$s} \\
    FPGA, initiation interval   & \multicolumn{4}{c}{$1.17\,\mu$s} \\
    \end{tabular}
    \end{ruledtabular}
\end{table}

\begin{table}[t!p]
    \footnotesize
    \caption{Per-component latency breakdown for the deployed stateful
        \texttt{hid128\_gru64\_dec128\_b1} kernel (Alveo U50,
        \texttt{io\_stream}/Resource strategy, $\mathrm{ReuseFactor}=64$, GRU hidden
        state as explicit kernel I/O). Cycle counts are Vitis HLS 2022.1 C-synthesis
        per-instance figures; wall-clock times use the real post-route achieved
        kernel clock of $104.7$~MHz ($9.55$~ns/cycle). The AXI wrapper rows are the
        physical interface cost of streaming the $117$ input and $118$ output words
        at one word per cycle; see the text for how to read the per-layer rows.}
    \label{tab:firmware-latency}
    \begin{ruledtabular}
    \scriptsize
    \begin{tabular}{lcc}
    \textrm{Component} & \textrm{Cycles} & \textrm{Time} \\
    \colrule
    AXI in: enqueue/demux (117 words) & $120$ & $1.146\,\mu$s \\
    Encoder L1 (Dense $53\!\to\!128$, 27-bit in) & $70$--$71$ & $0.669$--$0.678\,\mu$s \\
    Encoder ReLU                                 & $0$ & $0$ \\
    Encoder L2 (Dense $128\!\to\!128$)           & $65$--$66$ & $0.621$--$0.630\,\mu$s \\
    BatchNorm layer                              & $63$ & $0.602\,\mu$s \\
    GRU (64-unit, stateful, $h$ explicit I/O)    & $70$--$71$ & $0.669$--$0.678\,\mu$s \\
    Concat (encoder $\oplus$ GRU)                & $0$ & $0$ \\
    Decoder B0 L1 (Dense $192\!\to\!128$)        & $65$--$66$ & $0.621$--$0.630\,\mu$s \\
    Decoder B0 ReLU                              & $0$ & $0$ \\
    Decoder B0 L2 (Dense $128\!\to\!128$)        & $65$--$66$ & $0.621$--$0.630\,\mu$s \\
    Decoder B1 L1 (Dense $128\!\to\!128$)        & $65$--$66$ & $0.621$--$0.630\,\mu$s \\
    Decoder B1 ReLU                              & $0$ & $0$ \\
    Decoder B1 L2 (Dense $128\!\to\!128$)        & $65$--$66$ & $0.621$--$0.630\,\mu$s \\
    Decoder B1 residual add                      & $0$ & $0$ \\
    Decoder B2 L1 (Dense $128\!\to\!128$)        & $65$--$66$ & $0.621$--$0.630\,\mu$s \\
    Decoder B2 ReLU                              & $0$ & $0$ \\
    Decoder B2 L2 (Dense $128\!\to\!128$)        & $65$--$66$ & $0.621$--$0.630\,\mu$s \\
    Mean head (Dense $128\!\to\!27$)             & $65$--$66$ & $0.621$--$0.630\,\mu$s \\
    LogVar head (Dense $128\!\to\!27$)           & $65$--$66$ & $0.621$--$0.630\,\mu$s \\
    Output concat (mean $\oplus$ logvar $\oplus$ $h_\mathrm{next}$) & $0$ & $0$ \\
    \textit{NN core subtotal (dataflow-overlapped)} & $158$--$159$ & $1.509$--$1.519\,\mu$s \\
    AXI out: dequeue (118 words)                 & $117$--$121$ & $1.117$--$1.156\,\mu$s \\
    \colrule
    Total end-to-end latency       & $276$ & $2.64\,\mu$s \\
    Initiation interval            & $122$ & $1.17\,\mu$s \\
    \end{tabular}
    \end{ruledtabular}
\end{table}

\section{Conclusions}

We have presented an end-to-end workflow for implementing a recurrent probabilistic dynamics model for tokamak plasma control on FPGA kernel. We detail the model changes and resulting accuracy, as well as the resource usage and latency of the final FPGA implementation. Starting from the architecture of Ref.~\cite{char2024fullshot}, a systematic size ablation showed that the parameter
budget can be cut by $94.9\%$ for a few percent of one-step accuracy degradation ($+1.9\%$ MSE /
$-2.7\%$ EV on the held-out test split), and 16-bit quantization-aware training adds a similarly small one-step cost ($+2.8\%$ MSE / $-4.8\%$ EV). Under deployment conditions, a free-running ensemble replay of a held-out discharge tracks the actuator-driven evolution of the full shot.

Carrying the reduced, quantized model through a complete \textsc{hls4ml}, Vitis HLS, and Vivado implementation, we obtained a real, loadable accelerator binary for Xilinx Alveo U50 device that fits comfortably within every FPGA resource budget and delivers deterministic single-timestep inference in $2.64\,\mu$, enabling online look ahead control methods.

These results show that a fast, uncertainty-aware, hardware-deployable plasma dynamics model is achievable on commodity FPGA hardware today, at a well-characterized and modest accuracy price. We hope this workflow prove useful for bringing learned recurrent neural network based models into real-time fusion control more broadly.

\begin{acknowledgments}
This material is based upon work supported by the U.S. Department of Energy, Office of Science, Office of Fusion Energy Sciences, using the DIII-D National Fusion Facility, a DOE Office of Science user facility, under Award DE-FC02-04ER54698. We thank the authors of Ref.~\cite{char2024fullshot} for their work, which this paper builds upon.

\textbf{Disclaimer.}~This report was prepared as an account of work sponsored by an agency of the United States Government. Neither the United States Government nor any agency thereof, nor any of their employees, makes any warranty, express or implied, or assumes any legal liability or responsibility for the accuracy, completeness, or usefulness of any information, apparatus, product, or process disclosed, or represents that its use would not infringe privately owned rights. Reference herein to any specific commercial product, process, or service by trade name, trademark, manufacturer, or otherwise does not necessarily constitute or imply its endorsement, recommendation, or favoring by the United States Government or any agency thereof. The views and opinions of authors expressed herein do not necessarily state or reflect those of the United States Government or any agency thereof.
\end{acknowledgments}

\appendix

\bibliography{ref}

\end{document}